\documentclass[11pt]{article}
\usepackage{moriond,epsfig}
\usepackage[lflt]{floatflt}

\def\Journal#1#2#3#4{{#1} {\bf #2}, #3 (#4)}

\def\NIM{\em Nucl. Instrum. Methods}

\def\NPB{{\em Nucl. Phys.} B}
\def\PLB{{\em Phys. Lett.}  B}
\def\PRL{\em Phys. Rev. Lett.}
\def\PRD{{\em Phys. Rev.} D}

\def\EPJC{{\em Eur. Phys. J.} C} 

\def\be{\begin{equation}}
\def\ee{\end{equation}}
\def\bea{\begin{eqnarray}}
\def\eea{\end{eqnarray}}

\begin{document}
\vspace*{4cm}
\title{STRUCTURE FUNCTIONS AND THE SPIN OF THE NUCLEON:\\
FROM HERMES TO COMPASS}

\author{F.H. HEINSIUS (for the HERMES and COMPASS collaborations)}

\address{Fakult\"at f\"ur Physik, Universit\"at Freiburg, \\
79104 Freiburg, Germany}

\maketitle\abstracts{
The HERMES and SMC experiments have determined the 
contribution of different
quark flavors to the nucleon spin in a large range of Bjorken-x
via semi-inclusive deep inelastic scattering.
The main
goal of the COMPASS experiment is to measure the gluon polarization in
the nucleon. In all experiments a polarized lepton is scattered off a
polarized nucleon. 
Latest results from HERMES and perspectives for COMPASS running in
2001 and beyond are presented. }

\section{Spin of the Nucleon}

The spin of the nucleon is known to be 1/2$\hbar$ 
(in the following: $\hbar$=1). 
However, the EMC-experiment has
found that the spin of the quarks contribute only by a small fraction to
the proton spin.
Ever since it has been a longstanding
problem how the nucleon spin is divided among the quarks and gluons.
All possible contributions are summarized in a sum rule,
which splits the total spin of the nucleon 
into the contributions from quarks,
$\Delta\Sigma = \Delta u + \Delta d + \Delta s +
\Delta \overline{u} + \Delta \overline{d} + \Delta \overline{s}$,
from gluons
$\Delta G$, and the orbital angular momentum of quarks and gluons, 
$L_q$ and $L_g$, respectively:
\begin{equation}
 \frac{1}{2}=\frac{1}{2}\Delta\Sigma+\Delta G+L_q+L_g
\end{equation}
The spin density contribution of the quarks to the nucleon 
spin $\Delta\Sigma$ can be probed in deep inelastic scattering. The 
latest result~\cite{hermes:deltaq} obtained by the HERMES collaboration is 
$\Delta\Sigma=0.30\pm0.04\pm 0.09$, clearly showing that the other
contributions to the nucleon spin are needed.

The gluon spin density $\Delta G$ can be probed in the photon-gluon fusion
process as planned in the COMPASS experiment.
It has been suggested that
the orbital angular momentum $L_q$ of the quarks can be probed
through a measurement of the total angular momentum of the quarks
$J_q = \frac{1}{2}\Delta\Sigma + L_q$ using the framework of 
generalized parton distributions.~\cite{Ji:DVCS}

Non-relativistic quark models, where no distinction is made between
constituent and current quarks, 
yield $\Delta\Sigma=1$. This is clearly incompatible with the data.
In bag models the quarks have (small) current-quark masses, 
so relativistic effects have to be taken into account.
This leads also to orbital contributions to the angular momentum.
Jaffe and Manohar estimated for this case a quark spin
contribution of about $\Delta\Sigma\approx 0.65 - 0.75$.~\cite{JaffeManohar}
Note that these models do not contain any gluon contributions.
A major step forward towards a QCD based understanding was reached in a
QCD sum rule approach by Balitski and Ji.~\cite{Ji:QCD}
By defining a gauge invariant definition of the
total angular momentum contributions from quarks and gluons in the 
nucleon they estimated:
$J_q=\frac{1}{2}\Delta\Sigma + L_q = 0.15$ and
$ J_g = \Delta G + L_g=0.35$.

Recently a full (non-quenched) lattice QCD calculation has succeeded to
evaluate the contributions of separated quark flavors to the nucleon 
spin~\cite{guesken}:
$\Delta u = 0.62(7), \Delta d =-0.29(6), \Delta s= -0.12(7)$. These numbers
are, within their fairly large errors, compatible with present data.
Another approach is based on the chiral soliton model (instanton model).
It predicts a flavor asymmetry of the polarized antiquark 
distribution.~\cite{dressler}
This awaits confirmation by the experiments.

Experimentally the polarization of the partons in the nucleon are probed in
spin-dependent deep inelastic scattering (DIS). In inclusive DIS the asymmetry
in the lepton nucleon scattering cross section is measured 
for the spin of the nucleon in the same and opposite 
direction to the lepton helicity:
\begin{equation}
 A_1(x)=\frac{\sigma^{1/2}-\sigma^{3/2}}{\sigma^{1/2}+\sigma^{3/2}}\approx
\frac{g_1(x,Q^2)}{F_1(x,Q^2)}
\end{equation}
with 
$F_1(x,Q^2)=\frac{1}{2}\sum_f e_f^2 q_f(x,Q^2)$ and $g_1(x,Q^2)=\frac{1}{2}\sum_f e_f^2 \Delta q_f(x,Q^2)$ 
being the unpolarized and polarized structure
function in the quark parton model, respectively,
$\Delta q_f(x,Q^2)=q_f^{\uparrow}(x,Q^2)-q_f^{\downarrow}(x,Q^2)$, $Q^2$ the negative
4-momentum transfer squared of the photon, 
and $x$ the Bjorken scaling variable.
In semi-inclusive DIS a hadron is detected in coincidence with the scattered
lepton. This allows to tag the flavor of the struck quark by the identification
of the fragmentation products. For example, a $\pi^+$ is predominantly
produced from an $u$ quark.

\section{HERMES results}

The HERMES experiment, located at the HERA storage ring at DESY, has
taken inclusive and semi-inclusive deep inelastic scattering data off
positrons on $^1$H, $^2$H and $^3$He targets.~\cite{hermes:NIM} 
The measured inclusive asymmetries~\cite{hermes:deltaq}, 
together with the asymmetries of
positive and negative hadrons, allow to extract the flavor
dependent polarization of the up quarks, down quarks and the sea quark
contribution (Fig.~\ref{fig:deltaq}).

\begin{floatingfigure}[l]{0.40\textwidth}
\centering\psfig{figure=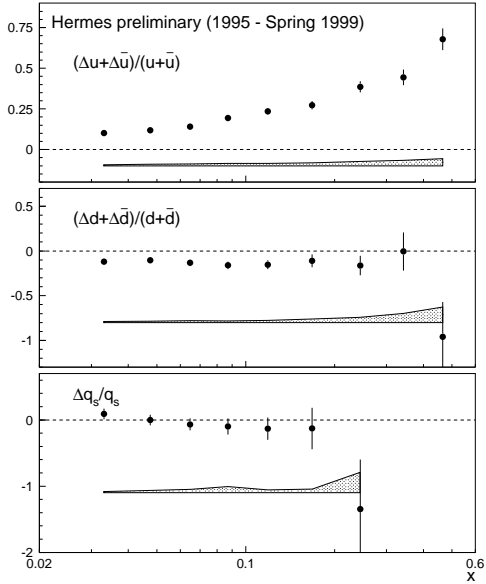,width=0.40\textwidth}
\caption{Polarization of up, down and sea quarks in the proton
as measured by the HERMES experiment. The sea polarization is assumed to
be flavor independent. All error bars shown are statistical errors, and the
bands indicate the systematic uncertainties.}
\label{fig:deltaq}
\end{floatingfigure}

The up quark polarization $(\Delta u + \Delta
\overline{u})/(u+\overline{u})$ is positive,
the down quark polarization $(\Delta d + \Delta
\overline{d})/(d+\overline{d})$ is slightly negative. 
Their absolute values are largest at large $x$ and remain different from
zero in the sea region. 
The
sea quark polarization $\Delta q_s /q_s$ is compatible with zero
in the measured range. In the extraction of the polarizations it was
assumed that the polarizations of the sea quarks are flavor independent.
New data taken on a deuterium target will allow a
much better constraint on the down quark polarization. Better
pion-kaon separation, due to the recently installed RICH detector,
will allow a flavor decomposition of the sea quarks.

From the inclusive deep inelastic scattering data the
structure function $g_1(x,Q^2)$ can be extracted. It is 
defined in the quark parton model as
\begin{equation}
g_1(x,Q^2)=\sum_{f=q,\overline{q}} e_f^2 \Delta q_f(x,Q^2).
\end{equation}
An analysis of data taken below $Q^2=1 $ GeV$^2$ extends the HERMES
$x$ range to $0.0021<x<0.021$.
The HERMES data~\cite{hermes:lowx} together with SMC data show a first hint on
scaling violations in  $g_1^p(x,Q^2)$, as are already well-known 
in the unpolarized structure function $F_1(x,Q^2)$.
These data are important to determine the quark spin contributions at small
$x$. This region has currently the largest contribution to the error in the 
determination of $\Delta\Sigma$.

The inclusive deep inelastic scattering data have also been used to attempt
a determination of the gluon spin contribution to the nucleon spin.
A QCD fit to the world data on the structure
function $g_1(x,Q^2)$ has been performed by 
the SMC collaboration.~\cite{smc:qcd} 
Through the evolution with $Q^2$ one gets access
to the singlet $\Delta \Sigma $ and non-singlet $\Delta q_{NS}$ 
quark contributions to the nucleon
spin as well as to the gluon contribution:
\begin{equation}
g_1(x,Q^2)=\frac{1}{9}(C_{NS} \otimes \Delta q_{NS} +
C_{\Sigma} \otimes \Delta \Sigma + 2 N_f C_g \otimes \Delta G )
\end{equation}
Here $C_{NS}$, $C_{\Sigma}$ and $C_g$ are 
splitting
 functions calculated 
in NLO QCD.
While the quark singlet spin contribution is determined very well 
($\Delta \Sigma = 0.38^{+0.03}_{-0.03}$$^{+0.03}_{-0.02}$$^{+0.03}_{-0.05}$),
this procedure does not put large constraints on the gluon
contribution to the spin 
($\Delta G = 0.99^{+1.17}_{-0.31}$$^{+0.42}_{-0.22}$$^{+1.43}_{-0.45}$ ).
A determination of the gluon spin contribution through evolution
requires measurements over a larger range of $Q^2$ and lower $x$
values. These will only be available at collider energies.
Therefore other possibilities are pursued in current experiments.

The first direct exploration of the gluon polarization has been
performed by the HERMES collaboration.~\cite{hermes:glue}
The analysis of high-$p_T$ hadron pairs
yielded a value of $\Delta G/G=0.41\pm 0.18$(stat) $\pm$ $0.03$(syst) at
an average fraction of the nucleon momentum carried by the struck gluon of
$<x_G>$=0.17. 
However, this determination is model dependent as a Monte
Carlo simulation is needed to determine the relative importance
of the photon-gluon fusion and the QCD Compton contributions to
the yield.

\section{COMPASS prospects}

\begin{figure}
\begin{center}
\psfig{figure=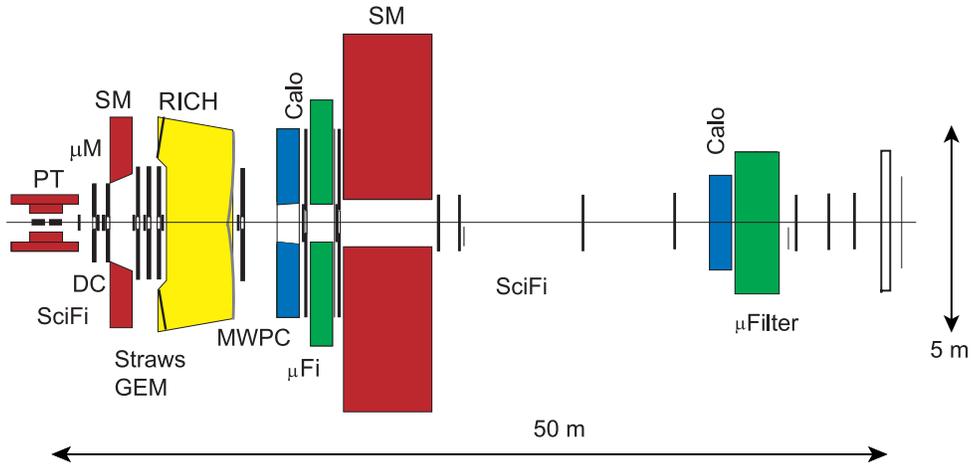,width=0.8\textwidth}
\end{center}
\caption{The COMPASS detector. Shown are the polarized target (PT), 
the two spectrometer magnets (SM), 
small angle tracking detectors (SciFi, $\mu$M, GEM),
the large angle tracking detectors (DC, Straws, MWPC) and
the particle identification detectors (RICH, hadron calorimeter $\mu$Filter).
\label{fig:compass}}
\end{figure}

The COMPASS~\cite{compass:proposal}  (Fig.~\ref{fig:compass})
experiment at CERN
consists out of a two-stage magnetic spectro\-meter with particle 
identification over a large kinematical range. 
A 100-200~GeV muon beam on $^6$LiD and NH$_3$ targets
will yield five times the luminosity compared to the SMC experiment.
High statistics measurements are made possible by the design of quasi
dead-time free readout electronics and event rates of up to
100~kHz.
The primary physics goals of the COMPASS experiment are the measurement
of the gluon polarization in the nucleon and the determination of the
longitudinal and transverse quark spin distribution functions for 
small $x$. In addition a rich program of hadron spectroscopy with hadron beams
is foreseen.

In COMPASS $\Delta G/G$
will be studied via the photon-gluon fusion process, tagged either by
open-charm production, or by high-$p_T$ hadron pair production.
The open-charm events will be identified by the reconstruction of D mesons
from the hadronic decay products. The measurement of the spin
dependent asymmetry for charm production allows the extraction of the gluon
polarization $\Delta G/G$ by unfolding the known photon-gluon cross section
and the gluon distribution $G(x)$. About 800 events with
reconstructed $D^0$ mesons are expected per day. 
This yields an uncertainty
of $\delta(\Delta G/G)=0.11$ in 2.5 years running time.
A high statistics measurement of high-$p_T$ hadron pairs allows
to measure several points of $\Delta G/G$ in the range of
$x_G=0.04-0.2$. However, the error is limited by systematics and is
expected to be $\delta(\Delta G/G)=0.05$ after one year running 
time.~\cite{bravar}       
COMPASS will start to take physics data in 2001. However, due to 
limitations in large angle tracking it is planned to start with the
measurement of $g_1$ at small $x$ values, which is important for the
study of possible scaling violations. In addition the experiment will
gather first data on the measurement
of $\Delta G/G$.

\section{Future}

The HERMES collaboration will continue the successful data taking in
run II (2001 -- 2006) with an emphasis on the measurement 
of the previously unknown
transverse quark spin distributions $h_1(x)$. 
In addition 
further
measurements of the
quark and gluon polarizations are planned. 
To get access to exclusive scattering a recoil detector around the
target is under development.

For the COMPASS experiment studies are ongoing to measure exclusive
scattering reactions like deeply virtual Compton scattering (DVCS).
Data on DVCS may provide information on the
        sofar unknown contribution of the orbital angular momentum of the
        quarks to the nucleon spin.
To ensure the exclusivity of these reactions a recoil detector is
required.                       


\end{document}